\documentclass[a4paper,11pt]{article}

\usepackage{jheppub}
\usepackage[T1]{fontenc}
\usepackage{hyperref}
\usepackage{graphicx}
\usepackage{amsmath}
\usepackage{epsfig}
\usepackage{subfig}
\usepackage{xcolor}
\usepackage{natbib}
\usepackage{multirow}
\usepackage[normalem]{ulem}

\title{Semi-analytical results for $e^+e^-\to J/\psi + X_{{\rm non\,}c\bar{c}}$ up to $\mathcal{O}(\alpha_s v^2)$ at B factories}
\author[a]{Cong Li,}
\author[a]{Hao-Yang Liu,}
\author[b]{Xu-Dong Huang,}
\author[a]{Wen-Long Sang}

\affiliation[a]{School of Physical Science and Technology, Southwest University, Chongqing 400700, China}
\affiliation[b]{College of Physics and Optoelectronic Engineering, Chongqing Normal University, Chongqing 401331, China}

\emailAdd{lc312321@163.com}
\emailAdd{tauceti@email.swu.edu.cn}
\emailAdd{huangxd@cqnu.edu.cn}
\emailAdd{wlsang@swu.edu.cn~(corresponding author)}

\abstract{
Within the NRQCD factorization framework, we investigate the color-singlet contribution to $e^+e^- \to J/\psi + X_{{\rm non\,}c\bar{c}}$ at B factories, computing the $\mathcal{O}(\alpha_s)$, $\mathcal{O}(v^2)$, and $\mathcal{O}(\alpha_s v^2)$ corrections to both the unpolarized cross section and the $J/\psi$ angular distribution. 
The $\mathcal{O}(\alpha_s v^2)$ correction is obtained for the first time, and the validity of NRQCD factorization at this order is explicitly verified.
Using the differential equation method, the short-distance coefficients are obtained as asymptotic expansions in $r = m_c/\sqrt{s}$ up to $r^{40}$, which reproduce exact results with high precision at B factory energies, achieving relative errors around $10^{-14}$ for the cross section and around $10^{-7}$ for the angular distribution.
Notably, with the same input parameters, our $\mathcal{O}(\alpha_s)$ and $\mathcal{O}(v^2)$ corrections are consistent with those reported in the literature.
Phenomenologically, the $\mathcal{O}(\alpha_s)$ correction (with $\mu_R=\sqrt{s}/2$) reaches about $50\%$ of the leading-order cross section, while the $\mathcal{O}(v^2)$ and $\mathcal{O}(\alpha_s v^2)$ corrections are accidentally small. After including feeddown contributions from $\psi(2S)$, the predicted cross section $0.523_{-0.197}^{+0.285}$ pb agrees with the {\tt Belle} measurement within uncertainties. However, the predicted angular distribution parameter $0.120_{-0.036}^{+0.041}$ deviates from the experimental value $5.71\pm 2.51$ by more than $2\sigma$, calling for further experimental and theoretical investigations.
}

\begin{document}

\maketitle

\bibliographystyle{JHEP}

\section{Introduction}
\label{sec:introduction}
Inclusive $J/\psi$ production at B factories provides an ideal testing ground for perturbative QCD and hadronization mechanisms. The first measurements of the cross section were reported by the {\tt BaBar}~\cite{BaBar:2001lfi} and {\tt Belle}~\cite{Belle:2002tfa,Belle:2001lqi,Belle:2009bxr} collaborations. The contributions from $c\bar{c}$ and non-$c\bar{c}$ final states were separately measured in Ref.~\cite{Belle:2009bxr}:
\begin{subequations}
\begin{eqnarray}
\label{cross:section}
\sigma(e^+e^-\to J/\psi+X_{c\bar{c}})&=&0.74\pm0.08^{+0.09}_{-0.08}~\text{pb},\\
\sigma(e^+e^-\to J/\psi+X_{{\rm non\,}c\bar{c}})&=&0.43\pm0.09\pm0.09~\text{pb},
\end{eqnarray}    
\end{subequations}
where the contributions from feeddowns of higher charmonium states are included.

Extensive theoretical studies have been performed~\cite{Keung:1980ev,Kiselev:1994pu,Braaten:1995ez,Yuan:1996ep,Cho:1996cg,Beneke:1997qw,Schuler:1998az,Liu:2003zr,Liu:2003jj,Fleming:2003gt,Zhang:2006ay,Ma:2008gq,Gong:2009ng,Gong:2009kp,He:2009uf,Jia:2009np,Zhang:2009ym,Shao:2014rwa,Lee:2020dza} (for more references, see the reviews~\cite{Brambilla:2010cs,Chen:2021tmf,Lansberg:2019adr}). We focus on the process $e^+e^-\to J/\psi+X_{{\rm non\,}c\bar{c}}$. The experimental measurement stands in stark contrast to the nonrelativistic QCD (NRQCD)  predictions~\cite{Keung:1980ev,Kuhn:1981jy,Kuhn:1981jn,Driesen:1993us,Cho:1996cg,Yuan:1996ep,Baek:1998yf,Hagiwara:2004pf} at leading order (LO) in both $\alpha_s$ and $v$, where $v$ denotes the typical relative velocity of the charm quark pair inside the $J/\psi$.~\footnote{Note that the NRQCD predictions at $v^0$ are identical to those in the color-singlet model.} 

Remarkably, the next-to-leading-order (NLO) QCD correction was computed in Refs.~\cite{Ma:2008gq,Gong:2009kp}, showing that the radiative correction leads to roughly a $20\%$ enhancement of the LO cross section. After including the QCD correction, the theoretical predictions for the cross section and the $J/\psi$ momentum distribution were found to agree with the experimental measurements; however, the $J/\psi$ angular distribution exhibits a significantly different shape compared to the experimental data~\cite{Gong:2009kp}.

The NLO relativistic correction was carried out in Refs.~\cite{He:2009uf,Jia:2009np}, which enhances the LO cross section by about $20-30\%$. Combining the relativistic correction with the QCD radiative correction, it has been argued that the color-singlet contribution alone can saturate the observed cross section for $e^+e^-\to J/\psi+X_{{\rm non\,}c\bar{c}}$, implying that the color-octet contributions are small~\cite{He:2009uf}.

QED contributions were also studied in Refs.~\cite{Zhang:2010uia,He:2013ka}, while the QED initial-state radiation effects were considered in Refs.~\cite{Shao:2014rwa,Gong:2019rpd}. Moreover, the color-octet contributions have been intensively investigated in Refs.~\cite{Braaten:1995ez,Yuan:1996ep,Beneke:1997qw,Fleming:2003gt,Wang:2003fw,Zhang:2009ym,Li:2014fya,Chen:2022qli}, and constraints on the color-octet matrix elements were performed in Refs.~\cite{Zhang:2009ym,Li:2014fya,Chen:2022qli}.

 In the present work, we restrict our attention to the color-singlet channel, leaving the color-octet contributions for future investigation.

Inspired by the considerable $\mathcal{O}(\alpha_s v^2)$ correction to the decay widths of $J/\psi\to 3\gamma$ and $J/\psi\to {\rm LH}$~\cite{Feng:2012by,Sang:2020zdv}, it is intriguing to consider this contribution for $e^+e^-\to J/\psi+X_{{\rm non\,}c\bar{c}}$ and to investigate its impact on the comparison between theory and experiment, particularly regarding the unresolved angular distribution puzzle. More importantly, NRQCD factorization for inclusive charmonium production has not yet been studied at $\mathcal{O}(\alpha_s v^2)$. Such a computation may serve as a stringent test of NRQCD factorization.

Furthermore, the NLO correction for $e^+e^-\to J/\psi+X_{{\rm non\,}c\bar{c}}$ has only been calculated numerically for specific choices of the charm quark mass and colliding energy $\sqrt{s}$. In phenomenology, it is desirable to obtain the analytical expressions, as they facilitate the study of the dependence of the theoretical computation on these parameters and enable the extraction of large logarithmic terms in the limit $\sqrt{s}\to \infty$. However, due to the complexity of multi-body phase-space integrals, deriving analytic expressions for this process is quite challenging. As an alternative, we derive a series expansion in $m_c/\sqrt{s}$, which includes various logarithmic terms of the form $\ln(m_c/\sqrt{s})$ and simultaneously provides a good approximation to the exact results. 

The structure of this paper is as follows. In Sec.~\ref{sec:kin:cross}, we analyze the kinematic structure and derive the generic formulas for the (differential) cross section in terms of two form factors. In Sec.~\ref{sec:NRQCD-factorization}, we present the NRQCD factorization for the form factors and outline the strategy for perturbative matching. In Sec.~\ref{sec:tech}, we describe the techniques used in our computation. In Sec.~\ref{sec:asy}, we examine the convergence of the asymptotic expansion and present the lowest-order expansion coefficients. Section~\ref{sec:phenomenology} is devoted to phenomenological analysis and discussion. Finally, we provide a summary in Sec.~\ref{sec:summary}.

\section{Kinematic structure and cross section~\label{sec:kin:cross}}

Let us consider the process $e^+(p_{e1})e^-(p_{e2})\to \gamma^*(k) \to J/\psi(P) + X$, where $X$ denotes light partons, i.e., gluons or light quarks. 
The cross section can be written generically as
\begin{equation}
\label{eq:dsigma-LH}
\sigma(e^+e^- \to J/\psi+X)
=\frac{1}{2s}\,\frac{1}{s^2}\,\frac{1}{4}\, \int L^{\mu \nu} H_{\mu \nu}\, d\Phi_{J/\psi}d\Phi,
\end{equation}
where the factor $1/4$ arises from the spin average over the initial state, and $1/s^2$ comes from the squared propagator of the virtual photon, with $s$ being the squared center-of-mass (CM) energy.
The phase space measures for the $J/\psi$ and light partons are given by
\begin{subequations}
\label{eq:phase-space}
\begin{eqnarray}
&&d\Phi_{J/\psi}=\frac{d^d P}{(2\pi)^{d-1}}\delta_+(P^2-M_{J/\psi}^2),\\
&&d\Phi=(2\pi)^d\delta^d(k-P-\sum_i k_i)\prod_i\frac{d^d k_i}{(2\pi)^{d-1}}\delta_+(k_i^2).
\end{eqnarray}
\end{subequations}
where $k_i$ denote the momenta of the light partons in the final state, 
and $d=4-2\epsilon$ denotes the spacetime dimensions. The function $\delta_+$ is the positive-energy delta function, defined as 
$\delta_+(p^2-m^2)\equiv \delta(p^2-m^2)\theta(p^0)$, which enforces both the on-shell condition and positive energy for the corresponding particle.

The leptonic tensor is given by
\begin{equation}
\label{eq:lepton-tensor}
 L^{\mu \nu}=4 \pi \alpha\, {\rm Tr}[\not{p}_{e1}\gamma^\mu\not{p}_{e2}\gamma^\nu],
\end{equation}
and the hadronic tensor $H_{\mu\nu}$, which represents the squared amplitude for $\gamma^*\to J/\psi+X$ with summation over final-state polarizations, is defined as
\begin{equation}
H_{\mu\nu} = \sum_{pol} \mathcal{M}_{\mu}\,\mathcal{M}_{\nu}^{*}.
\end{equation}
Integrating $H_{\mu\nu}$ over all the final states except the $J/\psi$ and using electromagnetic current conservation, we obtain
\begin{equation}
\label{eq:hadron-tensor}
\int H_{\mu \nu}\,d\Phi=H_1\left(-g_{\mu\nu}+\frac{k_\mu k_\nu}{k^2}\right)+H_2\left(P_\mu-\frac{P\cdot k}{k^2}k_\mu\right)\left(P_\nu-\frac{P\cdot k}{k^2}k_\nu\right),
\end{equation}
where the form factors $H_1$ and $H_2$ depend on scalar products of $P$ and $k$, and are independent of the momenta of the initial lepton pair.

Substituting Eqs.~\eqref{eq:lepton-tensor} and \eqref{eq:hadron-tensor} into Eq.~\eqref{eq:dsigma-LH} yields
\begin{eqnarray}
\label{eq:dsigma-LH-1}
&&\sigma(e^+e^- \to J/\psi+X)
= \frac{1}{2s}\,\frac{1}{s^2}\,\frac{1}{4}\, 4\pi\alpha\, s \int \bigg[4 H_1+2|\mathbf{P}|^2(1-\cos^2\theta) H_2\bigg] d\Phi_{J/\psi}\nonumber\\
&=&\frac{1}{2s}\,\frac{1}{s^2}\,\frac{1}{4}\, 4\pi \alpha\, s \int \bigg[-g^{\mu\nu}H_{\mu\nu}
  (1+\cos^2\theta) +\frac{P^{\mu} P^{\nu}}{|\mathbf{P}|^2}H_{\mu\nu} (1-3\cos^2\theta)
\bigg]d\Phi_{J/\psi}d\Phi,
\end{eqnarray}
where $\theta$ is the polar
angle between the outgoing $J/\psi$ and the electron beam direction, and $\mathbf{P}$ denotes the three-momentum of the $J/\psi$ in the CM frame. 
Note that in the second equality of Eq.~\eqref{eq:dsigma-LH-1}, we have reexpressed $H_1$ and $H_2$ in terms of $H_{\mu\nu}$ by solving Eq.~\eqref{eq:hadron-tensor}.

Since both $-g^{\mu\nu}H_{\mu\nu}$ and $P^{\mu} P^{\nu}H_{\mu\nu}$ are independent of $\theta$, we can write 
\begin{eqnarray}
\label{eq:dsigma-LH-2}
&&\frac{d\sigma(e^+e^- \to J/\psi+X)}{d\cos\theta}
=\frac{\pi\alpha^2 \alpha_{s}^{2}}{2s^2}\,  \bigg[A_1
(1+\cos^2\theta) +A_2 (1-3\cos^2\theta)
\bigg],
\end{eqnarray}
with
\begin{eqnarray}
\label{eq:Ai}
A_1=\frac{1}{2 \alpha \alpha_s^{2}}\int \bigg(-g^{\mu\nu}H_{\mu\nu}\bigg)d\Phi_{J/\psi}d\Phi,\quad
A_2=\frac{1}{2 \alpha \alpha_s^{2}}\int\bigg(\frac{P^{\mu} P^{\nu}}{|\mathbf{P}|^2}H_{\mu\nu}\bigg)d\Phi_{J/\psi}d\Phi.
\end{eqnarray}

The total cross section is obtained by integrating Eq.~\eqref{eq:dsigma-LH-2} over $\cos\theta$ from $-1$ to $1$:
\begin{eqnarray}
\label{eq:sigma}
\sigma(e^+e^- \to J/\psi+X)
=\frac{4\pi\alpha^2  \alpha_s^{2}}{3s^2}A_1.
\end{eqnarray}
Note that only $A_1$ contributes to the integrated cross section, while both $A_1$ and $A_2$ contribute to the angular distribution.

Thus, our main task reduces to the computation of $A_1$ and $A_2$.

\section{NRQCD factorization}
\label{sec:NRQCD-factorization}
\subsection{NRQCD factorization for $A_i$}

Within the NRQCD factorization framework~\cite{Bodwin:1994jh}, the quantities $A_i$ (with $i=1,2$) can be expressed as
\begin{equation}
\label{eq:nrqcd:Ai}
A_i =2M_{J/\psi}\langle \mathcal{O}(^3S_1) \rangle_{J/\psi} \bigg[F_i+G_i\langle v^2 \rangle_{J/\psi}+\mathcal{O}(v^4)\bigg],
\end{equation}
where $F_i$ and $G_i$ correspond to the short-distance coefficients (SDCs) associated with the $\mathcal{O}(v^0)$ and $\mathcal{O}(v^2)$ long-distance matrix elements (LDMEs), respectively, and the prefactor $2M_{J/\psi}$ appears because we employ the relativistic normalization for the $J/\psi$ state
in $A_i$, while the LDMEs adopt the
nonrelativistic normalization. The color-singlet LDMEs are defined as
\begin{subequations}
\label{eq:ldmes}
\begin{eqnarray}
\langle \mathcal{O}(^3S_1) \rangle_{J/\psi}&= &\frac{1}{3}\langle 0|\chi^{\dagger}\boldsymbol{\sigma}\psi \sum_{X,\,pol}|
J/\psi+X \rangle \cdot\langle J/\psi+X| \psi^{\dagger}\boldsymbol{\sigma}\chi|0\rangle, \\
\langle v^2 \rangle_{J/\psi} &=&
\frac{\langle \mathcal{P}(^3S_1) \rangle_{J/\psi}}
{m_c^2 \langle \mathcal{O}(^3S_1) \rangle_{J/\psi}},
\end{eqnarray}    
\end{subequations}
where $\psi^{\dagger}$ and $\chi$ are the Pauli spinor fields that create a charm quark and an anticharm quark, respectively, and
\begin{equation}
\mathcal{P}(^3S_1) = \frac{1}{6}\Big[
 \langle 0|\chi^{\dagger}\boldsymbol{\sigma}\psi
 \sum_{X,\,pol}|
J/\psi+X \rangle \cdot\langle J/\psi+X| \psi^{\dagger}\boldsymbol{\sigma}
 (-\tfrac{i}{2}\overleftrightarrow{\mathbf{D}})^{2}\chi|0\rangle
 + \mathrm{H.c.} \Big],
\end{equation}
where $\overleftrightarrow{\mathbf{D}}$ is defined by 
$\chi^\dagger \overleftrightarrow{\mathbf{D}} \psi = \chi^\dagger (\mathbf{D} \psi) - (\mathbf{D} \chi)^\dagger \psi$ with $D^\mu=\partial^\mu+igA^\mu$.  

The SDCs $F_i$ and $G_i$ are insensitive to long-distance nonperturbative physics and are therefore perturbatively calculable. 

\subsection{Determine the SDCs}
\label{sec:sdcs}
According to the factorization spirit, the SDCs are insensitive to the hadronization effects of the $J/\psi$. Hence, we may replace the physical $J/\psi$ state by a free $c\bar{c}$ pair carrying the same quantum numbers (i.e., $^3S_1$). Eq.~\eqref{eq:nrqcd:Ai} then becomes 
\begin{equation}
\label{eq:matching}
A_i(c\bar{c}(^3S_1)) =\langle \mathcal{O}(^3S_1) \rangle_{c\bar{c}(^3S_1)} \bigg[F_i+G_i\langle v^2 \rangle_{c\bar{c}(^3S_1)}\bigg],
\end{equation}
where the prefactor $2M_{J/\psi}$
has been omitted because we adopt the relativistic
normalization for the charm quark states on both sides.
Both sides can now be computed perturbatively and expanded in $\alpha_s$ and $v^2$. The SDCs $F_i$ and $G_i$ can then be determined order by order in $\alpha_s$. 

We first outline the computation of the left-hand side of Eq.~\eqref{eq:matching}. The amplitude for $\gamma^*(k)\to c(p)\bar{c}(\bar{p})+X$ is generated, where at LO and NLO in $\alpha_s$, $X$ denotes $gg$ and $ggg\,(gq\bar{q})$, respectively. The momenta $p$ and $\bar{p}$ are defined as
\begin{equation}
p = \frac{P}{2} + q, \qquad
\bar{p} = \frac{P}{2} - q,
\end{equation}
where $P$ and $q$ denote the total momentum and half of the relative momentum  of the charm pair, respectively. The on-shell conditions imply
\begin{equation}
P \cdot q = 0, \qquad
P^2 = 4 E^2,
\end{equation}
where $E = \sqrt{m_c^2 - q^2}$. Note that $q^2=-\mathbf{q}^2$. 

To project the $c\bar{c}$ pair onto the desired color-singlet spin-triplet state, we employ the relativistically normalized covariant projector~\cite{Bodwin:2002cfe}:
\begin{equation}
\label{eq:projector}
\Pi
= \frac{1}{4\sqrt{2}\, E (E + m_c)}
(\not{\bar{p}} - m_c)\not{\epsilon}^{*}
(\not{P} + 2 E)(\not{p} + m_c)
\otimes \frac{\mathbf{1}_c}{\sqrt{N_c}},
\end{equation}
where $\epsilon$ is the spin-1 polarization vector  satisfying $P\cdot\epsilon=0$.
The amplitude $\mathcal{M}^\mu$ is then obtained as
\begin{equation}
\label{eq:amp:spin1}
\mathcal{M}^\mu={\rm Tr}[\widetilde{\mathcal{M}}^\mu\;\Pi],
\end{equation}
where $\widetilde{\mathcal{M}}^\mu$ denotes the amplitude for $\gamma^*\to c\bar{c}+X$ with the external charm spinors truncated. 

We next project out the S-wave component from \eqref{eq:amp:spin1}. A technical subtlety
deserves mention. First, part of the relativistic correction is encoded in the phase space integral, because the invariant mass of the $c\bar{c}$ pair is $P^2=4E^2$ rather than $4m_c^2$.
Second, the amplitude involves $P\cdot k_i$, which depend on $E$ rather than $m_c$. As
pointed out in Refs.~\cite{Jia:2009np,Sang:2020zdv}, it is convenient to first replace $m_c$ with $\sqrt{E^2+q^2}$ and expand
the amplitude in powers of $q^2/E^2$. Finally, after carrying out the phase space integrals, we revert to more conventional expressions by expanding
$E=\sqrt{m_c^2-q^2}$ in powers of $q^2/m_c^2$, which automatically accounts for the relativistic effects in the phase space.

To this end, we first substitute $m_c$ to $\sqrt{E^2+q^2}$ and expand 
$\mathcal{M}^\mu$ up to second order in $q^2/E^2$
\begin{equation}
\label{eq:swave}
\mathcal{M}^\mu=\mathcal{M}_{0}^\mu+\mathcal{M}_{2}^\mu+\mathcal{O}(q^4/E^4),
\end{equation}
where
\begin{subequations}
\label{eq:S-wave-amplitudes}
\begin{align}
\mathcal{M}_{0}^{\mu}&=\mathcal{M}^{\mu}\big|_{\mathbf{q} \to 0},\\
\mathcal{M}_{2}^{\mu}&=\frac{\mathbf{q}^2}{d-1}\mathcal{I}^{\rho \beta}\,\frac{1}{2!}\,
\frac{\partial^2 \mathcal{M}^\mu}{\partial q^{\rho}\partial q^{\beta}}\bigg|_{\mathbf{q}\to 0},
\end{align}    
\end{subequations}
where $\mathcal{I}^{\rho \beta}=-g^{\rho \beta}+\frac{P^\rho P^\beta}{ P^2}$.

The hadronic tensor $H_{\mu\nu}$ is obtained by squaring the amplitude and summing over final-state polarizations. The quantities $A_i$ defined in \eqref{eq:Ai} can then be computed by integrating over the phase space. Expanding $A_i$ in powers of $q^2/E^2$ and $\alpha_s$ yields
\begin{eqnarray}
\label{eq:Ai:pert-1}
A_i&=&a_i^{(0)}(E^2,s)+\frac{\alpha_s}{\pi} a_i^{(1)}(E^2,s)+\frac{\mathbf{q}^2}{E^2} \bigg[b_i^{(0)}(E^2,s)+\frac{\alpha_s}{\pi}b_i^{(1)}(E^2,s)\bigg],
\end{eqnarray}
where $a_i$ and $b_i$ will be computed as an asymptotic expansion in powers of $\frac{E}{\sqrt{s}}$. The techniques used to compute $a_i$ and $b_i$ will be presented in Sec.~\ref{sec:tech}. A detailed comparison between the asymptotic expansion and the exact numerical results for $a_i$ and $b_i$ will be presented in Sec.~\ref{sec:comparison}.

Once $a_i$ and $b_i$ are obtained,
it is straightforward to expand $E$ in powers of $q^2/m_c^2$
\begin{eqnarray}
\label{eq:Ai:pert-2}
A_i&=&a_i^{(0)}(m_c^2,s)+\frac{\alpha_s}{\pi} a_i^{(1)}(m_c^2,s)+v^2 \bigg[\bigg(m_c^2\frac{\partial a_i^{(0)}(m_c^2,s)}{\partial m_c^2}+b_i^{(0)}(m_c^2,s)\bigg)\nonumber\\
&&+\frac{\alpha_s}{\pi}\bigg(m_c^2\frac{\partial a_i^{(1)}(m_c^2,s)}{\partial m_c^2}+b_i^{(1)}(m_c^2,s)\bigg)\bigg]+\mathcal{O}(v^4),
\end{eqnarray}
where $v^2=\frac{\mathbf{q}^2}{m_c^2}$.

On the other hand, the perturbative NRQCD matrix elements are given by~\cite{Luke:1997ys,Bodwin:2008vp,Dong:2012xx,Feng:2012by,Li:2013otv,Xu:2014zra,Chung:2020zqc}:
\begin{subequations}
\label{eq:matrix-elements-pert}
\begin{eqnarray}
\langle \mathcal{O}(^3S_1) \rangle_{c\bar{c}(^3S_1),\overline{\rm MS}} &=& 2 N_c (2E)^2\bigg[1+v^2\frac{\alpha_s}{\pi}\frac{16}{9\epsilon}\bigg]
=2 N_c (2m_c)^2\bigg[1+v^2\bigg(1+\frac{\alpha_s}{\pi}\frac{16}{9\epsilon}\bigg)\bigg],\phantom{xxxx}\\
\langle \mathcal{P}(^3S_1) \rangle_{c\bar{c}(^3S_1)}
 &=& 2 N_c \mathbf{q}^2 (2E)^2=2 N_c \mathbf{q}^2 (2m_c)^2(1+v^2),\\
\langle v^2 \rangle_{c\bar{c}(^3S_1)}
 &=& v^2,
\end{eqnarray}    
\end{subequations}
where we have adopted the $\overline{\rm MS}$ scheme to remove the UV divergence, and truncated the expansion up to $\mathcal{O}(v^2)$. 

Substituting Eqs.~\eqref{eq:Ai:pert-2} and \eqref{eq:matrix-elements-pert}
into Eq.~\eqref{eq:matching} allows us to determine the SDCs $F_i$ and $G_i$ order by order in $\alpha_s$.

Finally, it is worth noting that part of the relativistic corrections is hidden in the prefactor $2M_{J/\psi}$ in Eq.~\eqref{eq:nrqcd:Ai}.
To isolate these relativistic corrections, we expand
\begin{eqnarray}
\label{eq:Mjpsi:expand}
M_{J/\psi}=2E=2\sqrt{m_c^2-q^2}=2m_c\bigg(1+\frac{1}{2}v^2\bigg)+\mathcal{O}(v^4),
\end{eqnarray}
which introduces additional relativistic corrections. Consequently, the SDCs associated with $\langle v^2 \rangle_{J/\psi}$ are modified accordingly. It is straightforward to incorporate these relativistic corrections by replacing $M_{J/\psi}$ with $2m_c$ and simultaneously replacing $G_i$ by $\frac{1}{2}F_i+G_i$ in  Eq.~\eqref{eq:nrqcd:Ai}.

\subsection{SDCs in powers of $\alpha_s$}
As mentioned, we can rewrite Eq.~\eqref{eq:nrqcd:Ai} as
\begin{eqnarray}
\label{eq:nrqcd-reexpand}
A_i &=& 4m_c\langle \mathcal{O}(^3S_1) \rangle_{J/\psi} \bigg[F_i+\bigg(\frac{1}{2} F_i+G_i\bigg)\langle v^2 \rangle_{J/\psi}+\mathcal{O}(v^4)\bigg]\nonumber\\
&=& 4m_c\langle \mathcal{O}(^3S_1) \rangle_{J/\psi} \bigg[F_i+\tilde{G}_i\langle v^2 \rangle_{J/\psi}+\mathcal{O}(v^4)\bigg].
\end{eqnarray}

In general, the SDCs can be written as 
\begin{subequations}
\label{eq:SDC_E}
\begin{align}
F_i(m_c^2,s)&=\frac{1}{2m_{c}^{2}}\Big[f_i^{(0)}(r)+\frac{\alpha_s}{\pi}\bigg(\frac{\beta_0}{2}\ln\frac{\mu_R^2}{m_c^2}\,f_i^{(0)}(r)+f_i^{(1)}(r)\bigg)\Big],\\
\tilde{G_i}(m_c^2,s)&=\frac{1}{2m_{c}^{2}}\Big[g_i^{(0)}(r)+\frac{\alpha_s}{\pi}\bigg(\frac{\beta_0}{2}\ln\frac{\mu_R^2}{m_c^2}\, g_i^{(0)}(r)
 +\frac{16}{9}\ln\frac{\mu_{\Lambda}^2}{m_c^2}\,f_i^{(0)}(r)+g_i^{(1)}(r)\bigg)\Big],
\end{align}    
\end{subequations}
where $r=m_c/\sqrt{s}$ and $\beta_0=\frac{11}{3}C_A-\frac{2}{3}n_f$ is the one-loop coefficient of the QCD $\beta$ function with $n_f=n_l+n_h$ denoting the number of active quark flavors. Here $n_l=3$ is the number of light quarks, and $n_h=1$ is the number of heavy quarks.
$\mu_R$ and $\mu_\Lambda$ denote the renormalization scale and the NRQCD factorization scale, respectively. The $\ln\mu_R^2$ term is required by renormalization group invariance, while the $\ln\mu_\Lambda$ term arises from the factorization procedure. Note that the $\mu_\Lambda$ dependence in $\tilde{G_i}$ is canceled by the corresponding dependence in $\langle \mathcal{O}(^3S_1) \rangle_{J/\psi}$.

\section{Computational techniques}\label{sec:tech}
\begin{figure}[h!]
\begin{center}
\includegraphics[width=0.80\textwidth]{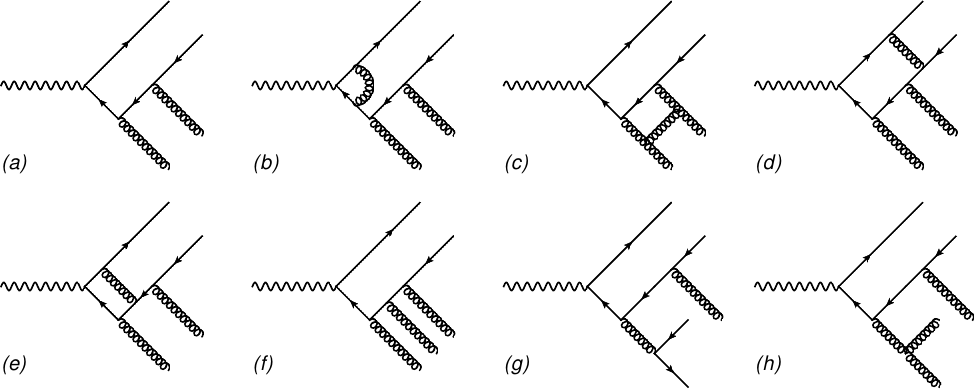}
\end{center}
\caption{Feynman diagrams for the process $\gamma^* \to c\bar{c}+X$. (a) is the LO diagram; (b)–(e) are virtual correction diagrams; (f)–(h) are real correction diagrams. The Feynman diagrams were drawn using {\tt JaxoDraw}.~\cite{Binosi:2008ig}.}
\label{fig:feynman}
\end{figure}

We use {\tt FeynArts}~\cite{Hahn:2000kx} to generate the Feynman diagrams and the corresponding amplitudes for $\gamma^*\to c\bar{c}+X$ up to NLO in $\alpha_s$. 
Some representative diagrams are shown in Fig.~\ref{fig:feynman}. We employ spin-triplet and color-singlet projectors to extract the spin-1 component, and apply Eq.~\eqref{eq:swave} to isolate the $S$-wave contribution of the $c\bar{c}$ pair. The squared amplitudes are then readily obtained. The Dirac algebra and Lorentz contractions are performed using {\tt FeynCalc} and {\tt FormLink}~\cite{Mertig:1990an,Feng:2012tk}.

Computing the three- and four-body phase space integrals directly turns out to be challenging. Alternatively, we adopt the {\it reverse unitarity method}, which converts phase space integrals into imaginary parts of multiloop integrals~\cite{Anastasiou:2002yz,Anastasiou:2003yy,Anastasiou:2002qz,Gehrmann-DeRidder:2003pne}. This allows us to employ standard multiloop computational techniques. 
Furthermore, the magnitude of the three-momentum $|\mathbf{P}|$ of the $J/\psi$ in the CM frame appears in Eq.~\eqref{eq:Ai}. This non-manifest Lorentz invariant complicates the use of tensor reduction via  IBP. Therefore, it is necessary to convert it as an explicitly Lorentz invariant quantity using the identity:
\begin{equation}
\frac{1}{|\mathbf{P}|^2}=\frac{\sqrt{s}}{2E}\Big[\frac{1}{(k+P)^2-(\sqrt{s}+2E)^2}-\frac{1}{(k+P)^2-(\sqrt{s}-2E)^2}\Big].
\end{equation}
It is worth noting that the denominators of the two propagators are positive, and thus insensitive to the $i\varepsilon$ prescription.

With these preparations, we reduce the Feynman integrals to a set of master integrals (MIs) using {\tt Blade}~\cite{Guan:2024byi} for the IBP reduction. After IBP reduction, we obtain about $4$ integral families with a total of $8$ master integrals at LO in $\alpha_s$, and about $114$ integral families with a total of $719$ master integrals at NLO in $\alpha_s$ (including both the virtual correction and real correction). 

We compute the MIs as a series expansion in $x\equiv E/\sqrt{s}$ using the differential equation (DE) method. 
For a complete set of MIs $\vec{J}$ within a given family, the DE take the form
\begin{equation}\label{eq-de}
    \frac{\partial}{\partial x}\vec{J}\;=\;M(x,\epsilon)\vec{J},
\end{equation}
where $M(x,\epsilon)$ is Fuchsian at $x=0$. Consequently, the MIs can be solved as an expansion of the form
\begin{equation}\label{eq-asy}
\vec{J}(x,\epsilon) = \sum_{\alpha} x^\alpha \sum_{k=0}^{k_\alpha}\sum_{n=0}^{\infty} \vec{a}_{\alpha,k,n}(\epsilon)\,x^{n}\ln^k x,
\end{equation}
with $\alpha_k$ being a finite set of characteristic exponents determined by the DE. Substituting Eq.~\eqref{eq-asy} into Eq.~\eqref{eq-de}, we determine the exponents $\alpha_k$ by matching the lowest-order terms and compute the coefficients $\vec{a}$ from recurrence relations and initial conditions. The initial conditions are obtained by using the package {\tt AMFlow}~\cite{Liu:2017jxz,Liu:2022chg,Liu:2022mfb,Liu:2020kpc}.
For further details, we refer the reader to  Refs.~\cite{Li:2025mng,Li:2025pbt,Liu:2026ray,Chen:2025qgy,Chen:2026maw}.

For renormalization, we adopt the on-shell scheme for the charm-quark mass and field strength, and the $\overline{\rm{MS}}$ scheme for the QCD coupling constant to remove UV divergences.

After carrying out the asymptotic expansions for all MIs, multiplying by the corresponding coefficients, and performing the renormalization procedure, we finally obtain the expressions for $a_i$ and $b_i$ (defined in Eq.~\eqref{eq:Ai:pert-1}) as asymptotic expansions in $x$.

A residual IR divergence remains in $b_i^{(1)}$; its coefficient is $\tfrac{16}{9}a_i^{(0)}$. Remarkably, this IR divergence can be exactly canceled by the IR divergences appearing in the perturbative NRQCD LDME $\langle \mathcal{O}(^3S_1) \rangle_{c\bar{c}(^3S_1)}$, as can be seen from Eq.~\eqref{eq:matrix-elements-pert}, thereby rendering the SDCs finite.
The finite SDCs $f^{(0)}_i$, $f^{(1)}_i$, $g_i^{(0)}$ and $g_i^{(1)}$ are then straightforwardly determined following the matching strategy outlined in Sec.~\ref{sec:sdcs}.

\section{Asymptotic expansions}
\label{sec:asy}
\subsection{Convergence behaviour of the asymptotic expansions for $a_i^{(1)}$ and $b_i^{(1)}$}
\label{sec:comparison}
In general, we can write the following expansions:
\begin{subequations}
\label{eq:SDC-series-x}
\begin{eqnarray}
a_{i}^{(1)}(x)&=&\sum_{m=0}^{\infty}\sum_{n=0}^{4}\tilde{a}^{(1)}_i(m,n)\,x^m \ln^n x,
\\
b_{i}^{(1)}(x)&=&\sum_{m=0}^{\infty}\sum_{n=0}^{4}\tilde{b}_i^{(1)}(m,n)\,x^m \ln^n x+\frac{16}{9\epsilon}a_{i}^{(0)}(x),
\end{eqnarray}
\end{subequations}
where $i=1, 2$ and $x=E/\sqrt{s}$. Note that the explicit IR divergence in $b_i^{(1)}$ has been extracted.

We now compare the asymptotic expressions with the exact numerical results for $a_i^{(1)}$ and $b_i^{(1)}$ to examine the convergence behaviour of the asymptotic expansions. 

It is convenient to introduce a shorthand notation:
\begin{subequations}
\begin{eqnarray}
\label{eq:SDC-series-trunc}
a_{i,k}^{(1)}(x)&=&\sum_{m=0}^{k}\sum_{n=0}^{4}\tilde{a}^{(1)}_i(m,n)\,x^m \ln^n x,
\\
b_{i,k}^{(1)}(x)&=&\sum_{m=0}^{k}\sum_{n=0}^{4}\tilde{b}_i^{(1)}(m,n)\,x^m \ln^n x,
\end{eqnarray}
\end{subequations}
where the subscript $k$ indicates that the series expansion in $x$ is truncated at $x^k$.  Furthermore, we denote the exact numerical results for $a_i$ by $a_{i,exact}^{(1)}$, and for $b_i$ with the explicit IR divergence removed by $b_{i,exact}^{(1)}$. We compute $a_{i,exact}^{(1)}$ and $b_{i,exact}^{(1)}$ at a set of $x$ values using the package {\tt AMFlow}.
To quantify the relative error between the asymptotic expansions truncated at $x^{40}$ and the exact results, we define
\begin{align}
\label{eq:delta}
\delta^{a}_{i,40}=\bigg\lvert\frac{a_{i,exact}^{(1)}-a_{i,40}^{(1)}}{a_{i,exact}^{(1)}}\bigg\rvert,
\qquad
\delta^{b}_{i,40}=\bigg\lvert\frac{b_{i,exact}^{(1)}-b_{i,40}^{(1)}}{b_{i,exact}^{(1)}}\bigg\rvert.
\end{align}

\begin{figure}[htbp]
    \centering
    \includegraphics[width=0.48\linewidth]{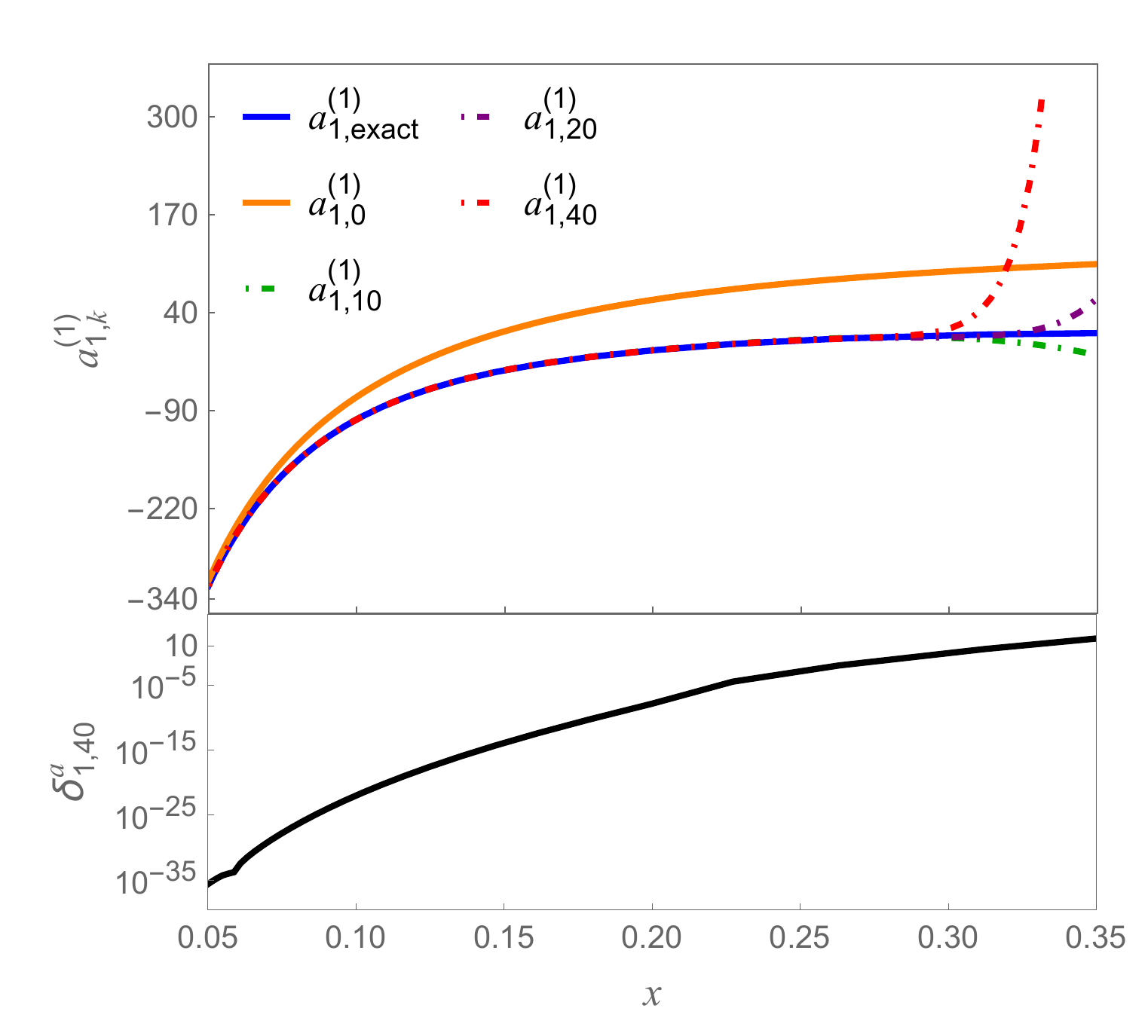}
    \hspace{0.02\textwidth}
    \includegraphics[width=0.48\linewidth]{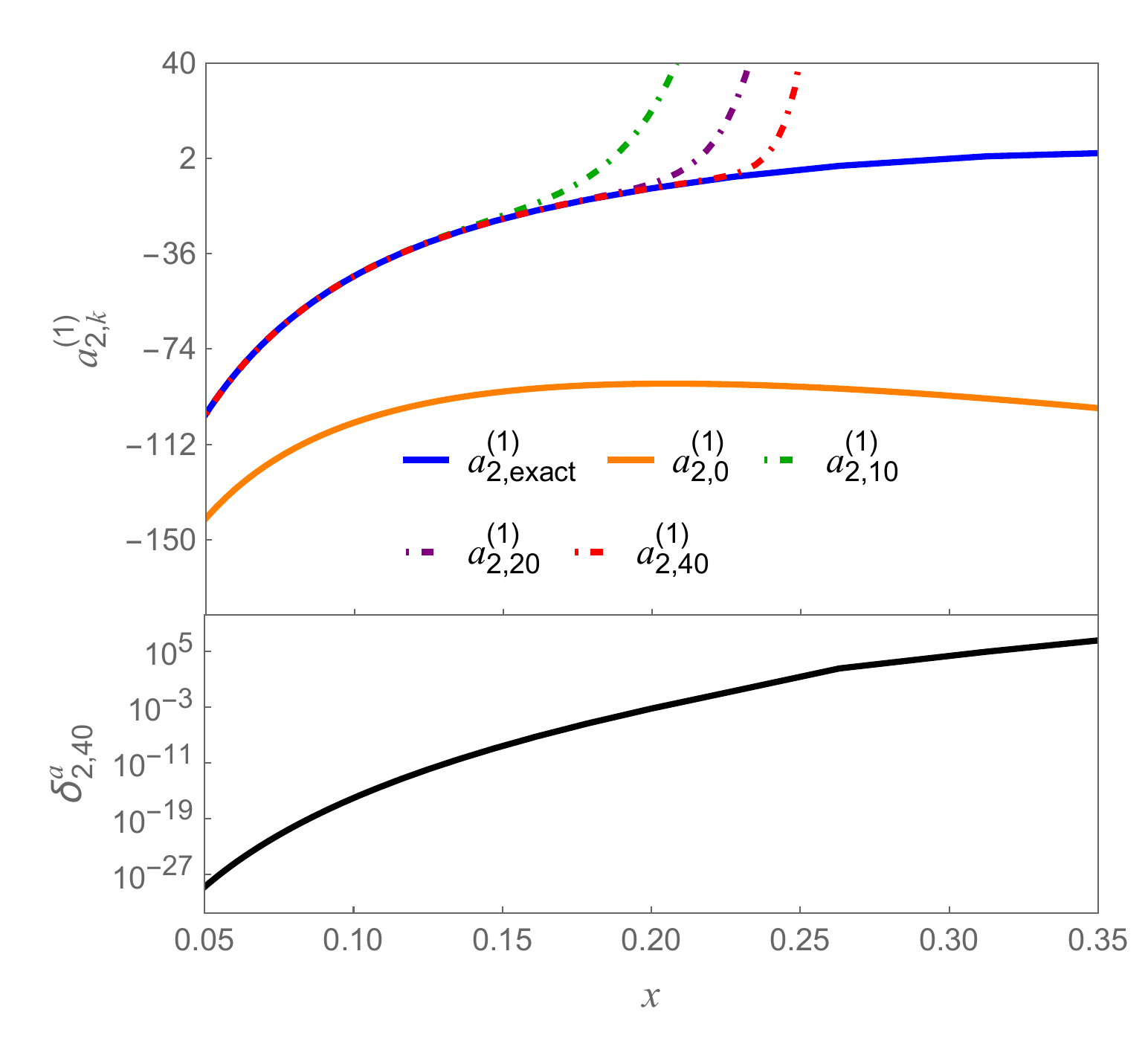}

    \vspace{0.3cm}

    \includegraphics[width=0.48\linewidth]{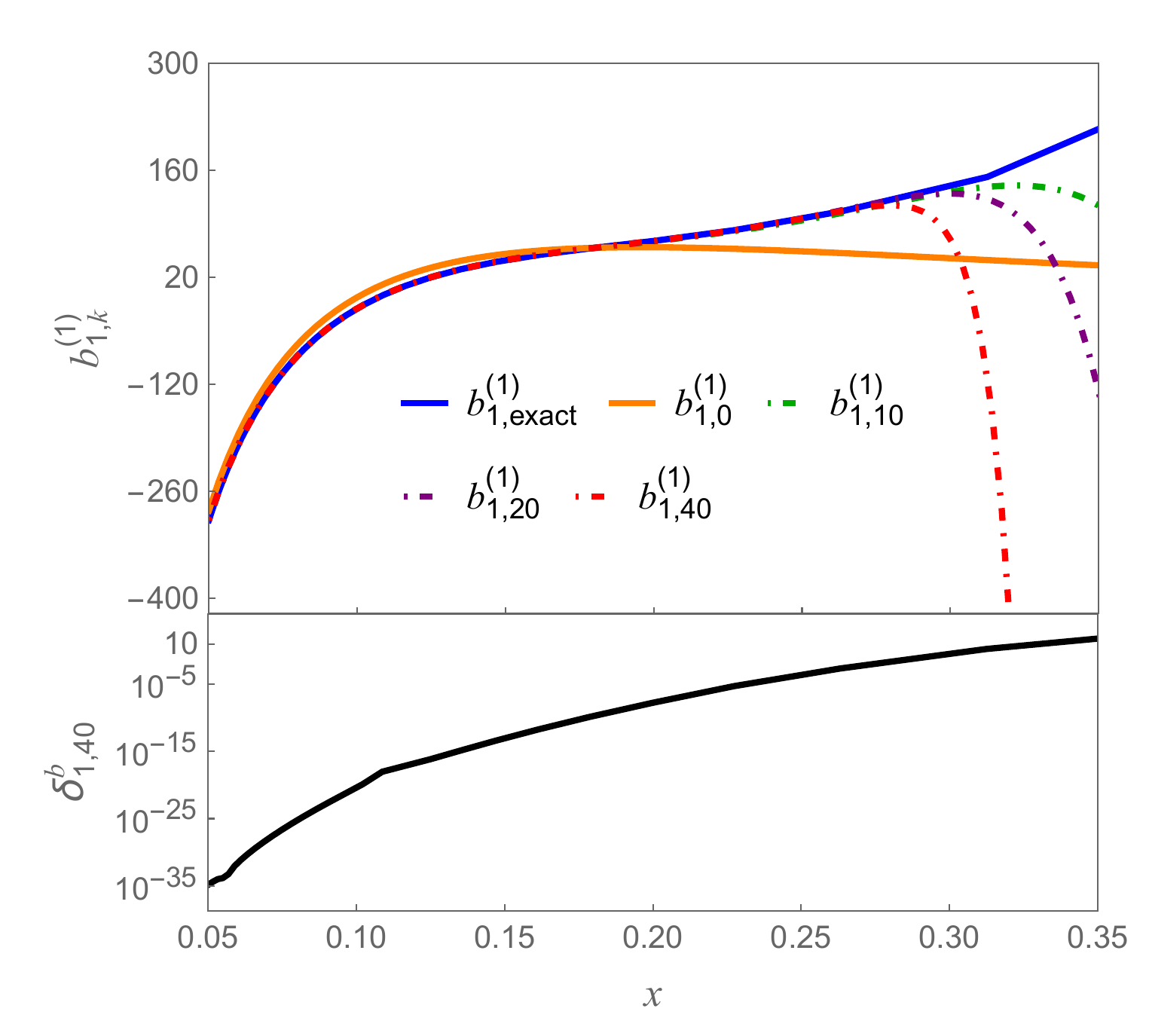}
    \hspace{0.02\textwidth}
    \includegraphics[width=0.48\linewidth]{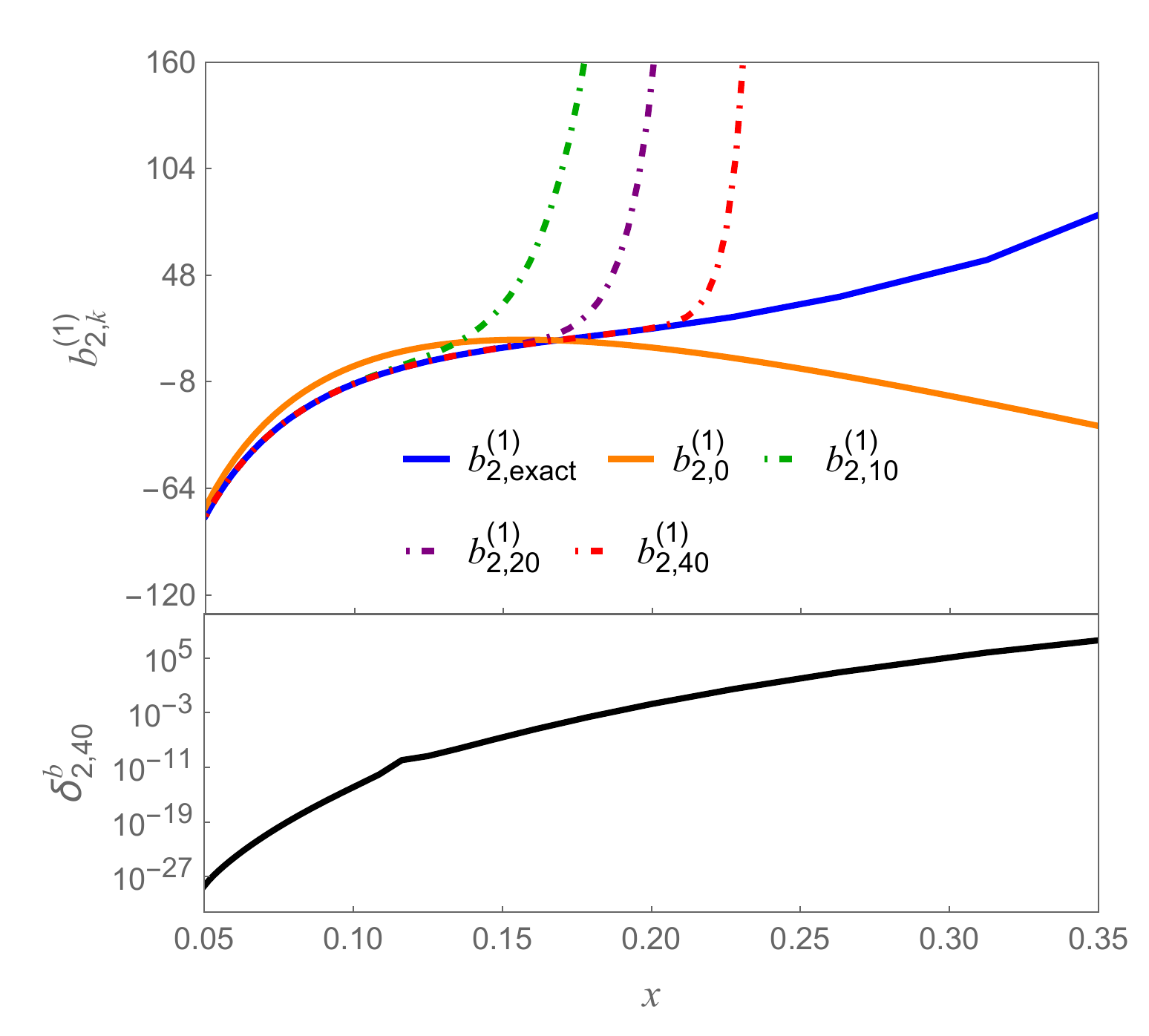}
    \caption{Comparison of the asymptotic expansion results with the exact numerical results.}
    \label{fig:asy:exa:comparison}
\end{figure}

The comparisons are shown in Fig.~\ref{fig:asy:exa:comparison}, where both the comparison between the asymptotic expansion results and the exact numerical values, and their relative errors, are presented. Several observations can be made.

The asymptotic expansion truncated at $x^0$ (orange line in the figure) approaches the exact results only at small $x$ and deviates significantly as $x$ increases. This is particularly evident for $a_2^{(1)}$, where the deviation is considerable even at $x=0.05$. Nevertheless, as the truncation order increases, the asymptotic expansion eventually converges to the exact results.

Taking $a_{1}^{(1)}$ as an example, the relative errors $\delta_{1,40}^{a}$ are $10^{-2}$, $1.4 \times 10^{-8}$, $10^{-15}$, and below $10^{-20}$ for $x=0.26$, $x=0.2$, $x=0.15$, and $x<0.1$, respectively. Similar behaviour is observed for 
$b_{1}^{(1)}$. Thus, the asymptotic expansions exhibit excellent convergence for $a_1^{(1)}$ and $b_1^{(1)}$ when $x<0.26$. 

Although the convergence for $a_2^{(1)}$ and $b_2^{(1)}$ is not as good as that for $a_1^{(1)}$, the relative errors for $a_2^{(1)}$ and $b_2^{(1)}$ remain about $10^{-4}$ and $10^{-2}$ at $x=0.2$. Hence, the asymptotic expansions also show good convergence for $a_2^{(1)}$ and $b_2^{(1)}$ at $x<0.2$.

For large values of $x$, specifically $x>0.3$ for $a_1^{(1)}$, $b_1^{(1)}$, and $x>0.23$ for $a_2^{(1)}$, $b_2^{(1)}$, the asymptotic expansions for all $a_i^{(1)}$ and $b_i^{(1)}$ deviate significantly from the exact results, indicating that the $x=0$ asymptotic expansion provides a poor approximation in these regions. Nevertheless, the asymptotic expansions obtained in this work are sufficiently accurate for phenomenological purposes.
Specifically, at $x=0.146$, which corresponds to $E=M_{J/\psi}/2$ and $\sqrt{s}=10.58$ GeV, the relative errors for $a_1^{(1)}$, $a_2^{(1)}$, $b_1^{(1)}$, and $b_2^{(1)}$ are about $10^{-15}$, $10^{-9}$, $10^{-14}$, and $10^{-7}$, respectively. 

\subsection{Asymptotic expressions for the SDCs}
\label{sec:asy:exp}
The SDCs $f_i^{(0)}$, $f_i^{(1)}$, $g_i^{(0)}$, and $g_i^{(1)}$
can be generically expressed as expansions of the form 
\begin{subequations}
\begin{align}
\label{eq:SDC-series-r}
f_i^{(0)}(r)=\sum_{m=0}^{\infty}\sum_{n=0}^{2}\tilde{f}^{(0)}_i(m,n)\,r^m \ln^n r,
\qquad
g_i^{(0)}(r)=\sum_{m=0}^{\infty}\sum_{n=0}^{2}\tilde{g}_i^{(0)}(m,n)\,r^m \ln^n r,\\
f_i^{(1)}(r)=\sum_{m=0}^{\infty}\sum_{n=0}^{4}\tilde{f}^{(1)}_i(m,n)\,r^m \ln^n r,
\qquad
g_i^{(1)}(r)=\sum_{m=0}^{\infty}\sum_{n=0}^{4}\tilde{g}_i^{(1)}(m,n)\,r^m \ln^n r,
\end{align}
\end{subequations}
where we have obtained high-precision numerical values for the expansion coefficients up to $r^{40}$. By applying the {\tt PSLQ} algorithm~\cite{ferguson1999analysis}, we are able to reconstruct their analytic expressions for the expansion coefficients at lower powers of $r$. 

The various expansion coefficients at LO in $r$ are given below
\begin{eqnarray}
\tilde{f}^{(0)}_{1}(0,2)&=&\frac{64}{81} ,~~\tilde{f}^{(0)}_{1}(0,1)=\frac{64}{81} ,~~\tilde{f}^{(0)}_{1}(0,0)=-\frac{32}{9} + \frac{320\ln 2}{81} ,\nonumber\\
\tilde{f}^{(0)}_{2}(0,1)&=&-\frac{64}{81},\qquad\tilde{f}^{(0)}_{2}(0,0)=-\frac{16}{81} - \frac{4\pi^2}{81} - \frac{64\ln 2}{81},
\end{eqnarray}

\begin{eqnarray}
\tilde{g}^{(0)}_{1}(0,2)&=&-\frac{32}{243},~~\tilde{g}^{(0)}_{1}(0,1)= \frac{736}{729},~~\tilde{g}^{(0)}_{1}(0,0)= \frac{944}{729} + \frac{1312}{729} \ln 2 ,\nonumber\\
\tilde{g}^{(0)}_{2}(0,1)&=&  \frac{32}{243} ,\qquad\tilde{g}^{(0)}_{2}(0,0)=-  \frac{8}{81} - \frac{10\pi^2}{243} + \frac{32}{243} \ln 2,
\end{eqnarray}

\begin{eqnarray}
\tilde{f}^{(1)}_{1}(0,4)&=&\frac{176}{729} ,\qquad\tilde{f}^{(1)}_{1}(0,3)=\frac{3008}{729} + \frac{512}{729}\ln 2 - \frac{256}{729}n_l  ,\nonumber\\
\tilde{f}^{(1)}_{1}(0,2)&=&\frac{4432}{243}  - \frac{272}{729}\pi^2 - \frac{1408}{243}\ln 2 - \frac{896}{729} n_l, \nonumber\\
\tilde{f}^{(1)}_{1}(0,1)&=& \frac{9064}{729} - \frac{5164}{2187}\pi^2  + \frac{17600}{729}\ln 2   - \frac{1072}{729}\pi^2\ln 2 \nonumber\\
&-& \frac{12352}{729}\ln^2 2 + \frac{4544}{243}\zeta(3) - \left(\frac{1664}{729}  - \frac{128}{729}\pi^2 \right)n_l  ,\nonumber\\
\tilde{f}^{(1)}_{1}(0,0)&=&12.9664 + 0.315077n_l,
\end{eqnarray}

\begin{eqnarray}
\tilde{f}^{(1)}_{2}(0,3)&=&- \frac{248}{729},~\quad
\tilde{f}^{(1)}_{2}(0,2)=-\frac{122}{27} - \frac{13}{486}\pi^2 - \frac{256}{243}\ln 2 + \frac{32}{81}n_l ,\nonumber\\
\tilde{f}^{(1)}_{2}(0,1)&=& -\frac{2660}{243} + \frac{221}{1458}\pi^2   - \frac{2432}{243}\ln 2 - \frac{31}{243}\pi^2\ln 2 - \frac{256}{243}\ln^2 2 \nonumber\\
&+& \frac{793}{486}\zeta(3) + \left(\frac{752}{729}  + \frac{4}{243} \pi^2 + \frac{128}{243} \ln 2\right)n_l,\nonumber\\
\tilde{f}^{(1)}_{2}(0,0)&=&  -19.7347 + 1.43278n_l ,
\end{eqnarray}

\begin{eqnarray}
\tilde{g}^{(1)}_{1}(0,4)&=&-\frac{88}{2187},\qquad
\tilde{g}^{(1)}_{1}(0,3)=\frac{256}{729} - \frac{256}{2187}\ln 2 +  \frac{128}{2187} n_l
,\nonumber\\
\tilde{g}^{(1)}_{1}(0,2)&=& \frac{3416}{2187} + \frac{248}{2187}\pi^2 + \frac{10432}{2187}\ln 2 + \frac{256}{729}\ln^2 2 -  \frac{448}{2187} n_l ,\nonumber\\
\tilde{g}^{(1)}_{1}(0,1)&=& \frac{71194}{6561} - \frac{250}{6561}\pi^2 + \frac{91360}{6561}\ln 2 - \frac{280}{2187}\pi^2\ln 2 \nonumber\\
&-& \frac{10976}{2187}\ln^2 2 - \frac{512}{2187}\ln^3 2 - \frac{472}{729}\zeta(3) \nonumber\\
&-& \left( \frac{2144}{6561} + \frac{32}{2187}\pi^2 + \frac{3584}{2187}\ln 2 \right)n_l ,\nonumber\\
\tilde{g}^{(1)}_{1}(0,0)&=& 13.2954 -  1.84618 n_l,
\end{eqnarray}

\begin{eqnarray}
\tilde{g}^{(1)}_{2}(0,3)&=&\frac{124}{2187} ,
~\quad\tilde{g}^{(1)}_{2}(0,2)= -\frac{53}{243} - \frac{65}{2916}\pi^2 + \frac{128}{729}\ln 2  - \frac{16}{243}n_l ,\nonumber\\
\tilde{g}^{(1)}_{2}(0,1)&=& -\frac{1657}{2187} - \frac{1141}{4374}\pi^2 - \frac{320}{243}\ln 2 + \frac{277}{1458}\pi^2\ln 2+ \frac{128}{729}\ln^2 2  \nonumber\\
&-& \frac{2515}{2916}\zeta(3)  - \left( \frac{152}{2187} - \frac{10}{729}\pi^2+ \frac{64}{729}\ln 2 \right)n_l ,\nonumber\\
\tilde{g}^{(1)}_{2}(0,0)&=& -3.61798 + 0.343606 n_l,
\end{eqnarray}
where we have retained the explicit $n_l$ dependence.
The complete expressions of the expansion coefficients up to $r^{40}$ are lengthy and are therefore provided in the attached electronic supplementary material files.

It is worth noting that the analytical expression for $f_1^{(0)}$ has already been established. Our expansion coefficients agree with those reported in Refs.~\cite{Gong:2009kp,Jia:2009np} when expanded as power series in $r$. The large logarithmic terms at $\mathcal{O}(\alpha_s)$ are obtained analytically, which can be used for further theoretical studies, e.g., for the resummation of large logarithms.

\section{Phenomenology}
\label{sec:phenomenology}
\subsection{Input parameters}
Prior to making phenomenological predictions for $e^+e^-\to J/\psi+X$, we specify the input parameters. The CM energy at $B$ factories is $\sqrt{s}=10.58$ GeV.  
The charm-quark pole mass is taken as $m_c=1.4\pm 0.2$ GeV, and the electromagnetic coupling constant is set to $\alpha(\sqrt{s})=1/130.9$.

The strong coupling constant $\alpha_s(\mu_R)$ is evaluated with $n_f=4$ active quark flavors. It is obtained by solving the renormalization group equation at two-loop accuracy using the package {\tt RunDec}~\cite{Herren:2017osy}. For the scale  $\mu_R=\sqrt{s}/2$, we obtain $\alpha_s(\sqrt{s}/2)=0.209$. To estimate the scale uncertainty, we vary the renormalization scale $\mu_R$ from $2m_c$ to $\sqrt{s}$, with the central value set at $\mu_R=\sqrt{s}/2$.
The NRQCD factorization scale is fixed at $\mu_\Lambda=1$~GeV.

The LDMEs are determined by fitting theoretical predictions, which include $\alpha_s$ correction and a class of velocity corrections, to the experimental data for the process $J/\psi\to e^+e^-$ in Ref.~\cite{Bodwin:2007fz}.
The resulting values are~\footnote{We have used the vacuum-saturation approximation to relate the production LDMEs to the corresponding decay LDMEs, with a relative error of $\mathcal{O}(v^4)$~\cite{Bodwin:1994jh}.}
\begin{align}
\label{eq:nrqcd:values}
\langle\mathcal{O}(^3S_1)\rangle_{J/\psi}
= 0.440^{+0.067}_{-0.055}~{\rm GeV}^3,\quad \langle v^2\rangle_{J/\psi}= 0.225^{+0.106}_{-0.088}.
\end{align}
It is worth noting that in deriving our theoretical formulas, we have converted the mass difference between $M_{J/\psi}$ and $2m_c$ into relativistic corrections using the Gremm–Kapustin relation~\cite{Gremm:1997dq}.
It is interesting to estimate the ambiguity arising from different choices of $\langle v^2\rangle_{J/\psi}$, i.e., comparing the value given in Eq.~\eqref{eq:nrqcd:values} with that computed from the Gremm–Kapustin relation~\cite{Gremm:1997dq}
\begin{equation}
\label{eq:GK:v2}
    \langle v^2 \rangle_{J/\psi}\approx\frac{M_{J/\psi}-2m_c}{m_c}=\frac{ 3.0969-2\times 1.4}{1.4}\approx 0.212,
\end{equation}
where the charm quark pole mass is fixed at its central value, the mass of the $J/\psi$ is taken from the Particle Data Group~\cite{ParticleDataGroup:2024cfk}.
The difference between this value and the central value in Eq.~\eqref{eq:nrqcd:values} is evidently very small. 
In practice, we adopt the values in Eq.~\eqref{eq:nrqcd:values} 
for our predictions.

\subsection{Cross section}
Using the asymptotic expansions derived in this work together with the input parameters listed above, we present the cross section for $e^+e^-\to J/\psi+X$ at $B$ factories in Tab.~\ref{tab:cross:section:direct} at various levels of accuracy.
We use the symbol ``vLO” to denote the sum of the LO contribution and the $\mathcal{O}(v^2)$ correction, and ``vNLO” to denote the sum of the NLO contribution together with the $\mathcal{O}(v^2)+\mathcal{O}(\alpha_s v^2)$ corrections. The quoted uncertainties arise from the variation of the renormalization scale, the charm quark mass, and the NRQCD LDMEs. As shown in Tab.~\ref{tab:cross:section:direct}, the renormalization scale uncertainties for the direct production cross section are approximately $55\%$, $55\%$, $22\%$, and $23\%$ for the LO, vLO, NLO, and vNLO, respectively. Meanwhile, the combined uncertainties from the charm quark mass and the NRQCD LDMEs are approximately $49\%$, $45\%$, $54\%$, and $49\%$ for the LO, vLO, NLO, and vNLO, respectively.  To clearly see the contributions from different orders, we list the individual contributions from various origins at two specific values of $\mu_R$ in Tab.~\ref{tab:cross:section:indiv}, with $m_c$ and the NRQCD LDMEs fixed at their central values.

\begin{table}[!h]
\begin{center}
\caption{ Direct production cross section (in pb) for $e^+e^-\to J/\psi+X$ at B factories. 
The first uncertainty is obtained by varying $\mu_R$ from $2m_c$ to $\sqrt{s}$, and the second uncertainty arises from the variations of the charm quark mass and the NRQCD LDMEs.
}
\label{tab:cross:section:direct}
\begin{tabular}{cccccc}
\hline
 & LO & vLO & NLO &  vNLO \\  \hline
$\sigma_{\rm direct}$ &$0.257_{-0.081-0.083}^{+0.141+0.127}$&$0.252_{-0.079-0.074}^{+0.138+0.114}$&$0.386_{-0.081-0.130}^{+0.086+0.208}$&$0.382_{-0.082-0.119}^{+0.088+0.189}$\\ \hline
\end{tabular}
\end{center}
\end{table}

\begin{table}[!h]
\begin{center}
\caption{Individual contributions to the direct production cross section (in pb) for $e^+e^-\to J/\psi+X$ at B factories from each piece, with $m_c$ and the NRQCD LDMEs fixed at their central values}.
\label{tab:cross:section:indiv}
\begin{tabular}{cccccc}
\hline
 & LO & $\mathcal{O}(v^2)$ & $\mathcal{O}(\alpha_s )$ &  $\mathcal{O}(\alpha_s v^2)$&  vNLO\\  \hline
$\sigma_{\rm direct}(\mu_R=2m_c)$ &$0.398$&$-0.008$&$0.074$&$0.007$&$0.470$\\ 
$\sigma_{\rm direct}(\mu_R=\frac{\sqrt{s}}{2})$ &$0.257$&$-0.005$&$0.129$&$0.002$&$0.382$\\ \hline
\end{tabular}
\end{center}
\end{table}

Several features can be observed.  The $\mathcal{O}(\alpha_s)$ correction is positive and substantial, increasing the LO prediction by a factor ranging from $19\%$ to $73\%$ depending on the choice of $\mu_R$. Including the radiative correction, we find that the uncertainty from $\mu_R$ becomes slightly milder.
It is worth noting that, with the same parameters, our $\mathcal{O}(\alpha_s)$ correction agrees with the result in Refs.~\cite{Ma:2008gq,Gong:2009kp}.

In contrast, both the $\mathcal{O}(v^2)$ and $\mathcal{O}(\alpha_s v^2)$ corrections are abnormally small, which is quite different from naive estimates of $\mathcal{O}(v^2)\sim 30\%$ and $\mathcal{O}(\alpha_s v^2)\sim 10\%$. 
This is attributed to the accidental smallness of the SDCs associated with $\langle v^2\rangle_{J/\psi}$.
Our $\mathcal{O}(v^2)$ correction is consistent with Ref.~\cite{Jia:2009np} after converting their results expressed in terms of the charmonium mass to the conventional results expressed in terms of the charm quark mass. The $\mathcal{O}(\alpha_s v^2)$ correction computed in this work is new. To the best of our knowledge, this is the first calculation of $\mathcal{O}(\alpha_s v^2)$ corrections for inclusive quarkonium production.

Finally, it is worth noting that the uncertainties arising from the charm quark mass and the NRQCD LDMEs exceed those from the $\mu_R$ variation and constitute the dominant sources of theoretical uncertainty in the final predictions.

It is worth noting that only the prompt production cross section for $e^+e^-\to J/\psi+X$ has been measured~\cite{Belle:2009bxr}, which includes both the direct $J/\psi$ production and feeddown contributions from higher charmonium states (mainly from $\psi(2S)$ and $\chi_{cJ}$), i.e., $e^+e^-\to \psi(2S)+X\to J/\psi+X$. As noted in Ref.~\cite{Ma:2008gq}, the feeddown contribution from $\chi_{cJ}$ is suppressed, therefore, we only need to consider the feeddown from $\psi(2S)$. The NRQCD matrix elements for $\psi(2S)$ are not well known. However, since the relativistic corrections are small (as shown in Tab.~\ref{tab:cross:section:indiv}), it is reasonable to follow the treatment of Ref.~\cite{Zhang:2006ay} to estimate the feeddown contribution from $\psi(2S)$ by simply multiplying the $J/\psi$ cross section by a constant factor:  
\begin{eqnarray}
\frac{\langle\mathcal{O}(^3S_1)\rangle_{\psi(2S)}}{\langle\mathcal{O}(^3S_1)\rangle_{J/\psi}} {\rm Br}(\psi(2S)\to J/\psi+X)&\approx&\frac{M_{\psi(2S)}^2\Gamma(\psi(2S)\to e^+e^-)}{M_{J/\psi}^2\Gamma(J/\psi\to e^+e^-)} {\rm Br}(\psi(2S)\to J/\psi+X)\nonumber\\
&=& 0.368\pm 0.017,
\end{eqnarray}
where the masses of the quarkonia, decay widths, and branching fraction are taken from the latest Particle Data Group~\cite{ParticleDataGroup:2024cfk}, and the uncertainty originates from these parameters. We have employed the relation $\langle\mathcal{O}(^3S_1)\rangle_{H}\propto M_H^2\Gamma(H\to e^+e^-)$~\cite{Zhang:2006ay}, where using $M_H$ instead of $2m_c$ may account for part of the relativistic effects. For comparison, this constant factor is estimated to be $0.29$ in Ref.~\cite{Gong:2009kp} and $0.355$ in Ref.~\cite{Zhang:2006ay,Ma:2008gq}.

\begin{table}[!h]
\begin{center}
\caption{Prompt cross section (in pb) for $e^+e^-\to J/\psi+X$ at B factories. The three theoretical uncertainties correspond to $\mu_R$ variation, the combined uncertainty from $m_c$ and the NRQCD LDMEs, and the feeddown contribution from $\psi(2S)$, respectively.}
\label{tab:cross:section:prompt}
\begin{tabular}{cccccc}
\hline
 & LO & vNLO & {\tt Belle}~\cite{Belle:2009bxr}\\ \hline
$\sigma_{\rm prompt}$ &$0.351_{-0.111-0.113-0.004}^{+0.192+0.174+0.004}$&$0.523_{-0.112-0.162-0.006}^{+0.120+0.258+0.006}$ & $0.43\pm 0.09\pm 0.09$\\ \hline
\end{tabular}
\end{center}
\end{table}

The theoretical prediction for the prompt cross section is presented in Tab.~\ref{tab:cross:section:prompt}, where the experimental result is also shown for comparison. It can be found in Tab.~\ref{tab:cross:section:prompt} that the uncertainty arising from the feeddown contribution of $\psi(2S)$ is approximately $1\%$, which is smaller than the uncertainties from other sources.
After including all corrections, the theoretical prediction is consistent with the {\tt Belle} measurement within the uncertainties.

\begin{figure*}[!h]
\begin{center}
\hspace{0cm}\includegraphics[width=0.8\textwidth]{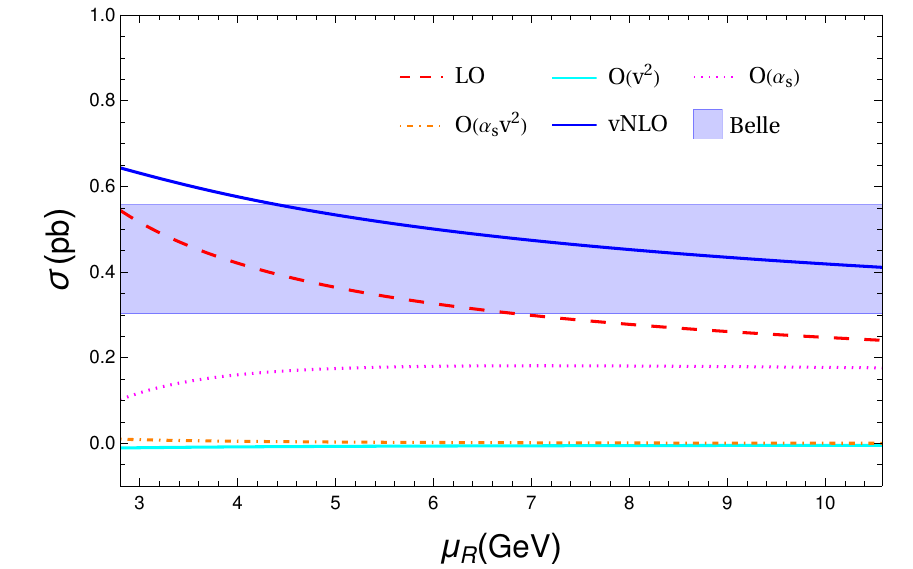}
\caption{\label{fig:cross:section}
Prompt cross section (in pb) for $e^+e^-\to J/\psi+X$ at $\sqrt{s} = 10.58$~GeV as a function of the scale $\mu_R$.  The other parameters are fixed at their central values.}
\end{center}
\end{figure*}

Additionally, to further examine the dependence of the theoretical prediction on the renormalization scale $\mu_R$, we show in Fig.~\ref{fig:cross:section} the prompt cross section from each component as a function of $\mu_R$ in the range from $2m_c$ to $\sqrt{s}$. It is observed that both the $\mathcal{O}(v^2)$ and $\mathcal{O}(\alpha_s v^2)$ corrections are quite small. Moreover,
we observe that the LO cross section decreases as $\mu_R$ increases, while the $\mathcal{O}(\alpha_s)$ correction increases with $\mu_R$. 
Consequently, the total cross section labeled by ``vNLO” exhibits a milder $\mu_R$ dependence than the LO. 

\begin{figure*}[!h]
\begin{center}
\hspace{0cm}\includegraphics[width=0.8\textwidth]{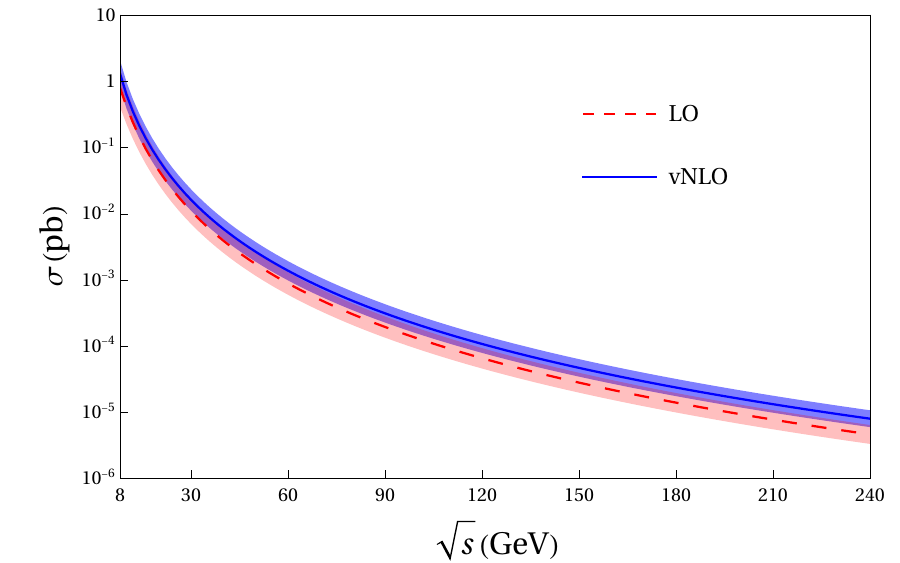}
\caption{\label{fig:s_run}
Prompt cross section (in pb) for $e^+e^-\to J/\psi+X$ as a function of the CM energy. The uncertainty is obtained by adding in quadrature the variations from $\mu_R$ (ranging from $\sqrt{s}/4$ to $\sqrt{s}$), the charm quark mass, the NRQCD LDMEs, and the feeddown contribution.}
\end{center}
\end{figure*}

Finally, it is interesting to investigate the cross section at different collision energies. We plot the prompt cross section as a function of $\sqrt{s}$ in Fig.~\ref{fig:s_run}. As $\sqrt{s}$ increases, the cross section decreases significantly, roughly as $1/s^2$.  
It is worth noting that as $\sqrt{s}$ approaches the $Z$ boson mass, the contribution from $e^+e^-\to Z^*\to J/\psi+X$, which is not computed here, becomes non-negligible and eventually dominant. However, as $\sqrt{s}$ moves away from the $Z$ peak, the cross section falls sharply. Consequently, observing this process at future Higgs factories would be very challenging.   

\subsection{Angular distribution}

\begin{figure*}[!h]
\begin{center}
\hspace{0cm}\includegraphics[width=0.8\textwidth]{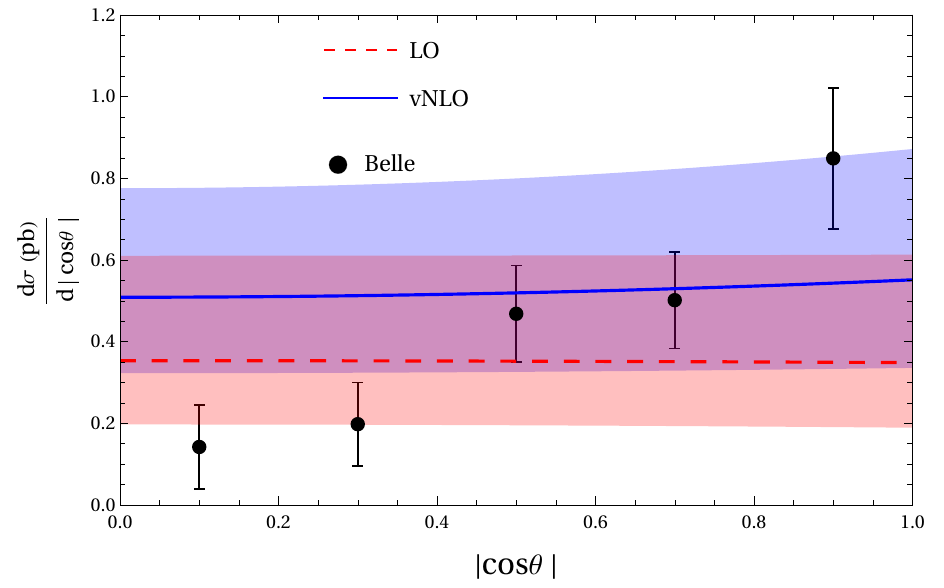}
\caption{\label{fig:angle}
Angular distribution for $e^+e^-\to J/\psi+X$ at $\sqrt{s} = 10.58$~GeV. The uncertainty is obtained by adding in quadrature the variations from $\mu_R$ (ranging from $2m_c$ to $\sqrt{s}$), the charm quark mass, the NRQCD LDMEs, and the feeddown contribution.}
\end{center}
\end{figure*}

Using Eq.~\eqref{eq:dsigma-LH-2}, we present the angular distributions for the prompt cross section of $e^+e^-\to J/\psi+X$ in Fig.~\ref{fig:angle}, where the experimental data from the {\tt Belle} collaboration~\cite{Belle:2009bxr} are also shown for comparison. Our theoretical predictions are in good agreement with the experimental data within uncertainties for the three data points on the right-hand side. For the two leftmost points, which correspond to small $\cos\theta$, the theoretical values are consistent with the experimental measurement within $2\sigma$.

It is also interesting to examine the line shape of the angular distribution. From Eq.~\eqref{eq:dsigma-LH-2}, we can derive
\begin{equation}
    \frac{d\sigma}{d\cos\theta}=\frac{\pi\alpha^2\alpha_{s}^{2}(A_1+A_2)}{2s^2}\bigg( 1+\alpha_\theta\cos^2\theta\bigg),
\end{equation}
where the angular distribution parameter is given by $\alpha_\theta=(A_1-3A_2)/(A_1+A_2)$. 
The theoretical prediction for $\alpha_\theta$ at different levels of accuracy can be straightforwardly obtained by expanding $\alpha_\theta$ in powers of $\alpha_s$ and $v^2$. It is evident that $\alpha_\theta$ is independent of the LO LDME. Moreover, as noted earlier, the relativistic corrections are small. Consequently, the theoretical prediction for $\alpha_\theta$ is insensitive to nonperturbative inputs, making $\alpha_\theta$ an ideal observable for testing NRQCD predictions. 

\begin{table}[!h]
\begin{center}
\caption{Angular distribution parameter $\alpha_\theta$ for $e^+e^-\to J/\psi+X$ at B factories. The source of the theoretical uncertainties is the same as in Tab.~\ref{tab:cross:section:direct}.}
\label{tab:ang:dis}
\begin{tabular}{cccccc}
\hline
 & LO & vNLO & {\tt Belle}~\cite{Belle:2009bxr}\\ \hline
$\alpha_\theta$ &$-0.014_{-0.000-0.031}^{+0.000+0.034}$ & $0.120_{-0.019-0.031}^{+0.027+0.031}$ &$5.71\pm 2.51$ \\ \hline
\end{tabular}
\end{center}
\end{table}

Our predictions for $\alpha_\theta$ are presented in Tab.~\ref{tab:ang:dis}. It can be found in Tab.~\ref{tab:ang:dis} that the combined uncertainties from $m_c$ and the NRQCD LDMEs of $\alpha_\theta$ are larger than those from the renormalization scale $\mu_R$. We also extract the experimental value of $\alpha_\theta$ through a least-squares fit of the measured angular distribution, and the results are listed in the table for comparison.   

One observation is that the $\mathcal{O}(\alpha_s)$ correction changes $\alpha_\theta$ from a small negative value to a small positive value. Consequently, although the magnitude of the differential cross section is significantly altered by the correction, the line shape of the angular distribution is only mildly affected, as can be seen in Fig.~\ref{fig:angle}.

It is evident that the theoretical prediction for $\alpha_\theta$ differs from the experimental data by more than $2\sigma$. In fact, a similar observation was made for the process $e^+e^-\to J/\psi+\chi_{c0}$~\cite{Liu:2026ray}, where the theoretical prediction for the cross section agrees with the experimental data, yet the angular distribution parameter shows a significant discrepancy. These puzzles certainly warrant further theoretical and experimental investigation.

\section{Summary}
\label{sec:summary}

Within the NRQCD factorization framework, we investigate the color-singlet contribution to the process $e^+e^-\to J/\psi + X$ at B factories, where $X$ denotes light hadronic final states.
Specifically, we compute the $\mathcal{O}(\alpha_s)$, $\mathcal{O}(v^2)$, and $\mathcal{O}(\alpha_s v^2)$ corrections to both the unpolarized cross section and the $J/\psi$ angular distribution. 
The $\mathcal{O}(\alpha_s v^2)$ correction is obtained for the first time, and the validity of NRQCD factorization at this order is explicitly verified.

Using the DE method, we obtain the SDCs as asymptotic expansions in $r= m_c/\sqrt{s}$ up to $r^{40}$, where the large logarithmic terms are presented analytically with the help of the {\tt PSLQ}  algorithm. 
These asymptotic expressions accurately reproduce the exact results for $r<0.26$ for the unpolarized cross section and for $r<0.20$ for the angular distribution. In particular, at $m_c=1.4$ GeV and $\sqrt{s}=10.58$ GeV, the relative errors are below $10^{-13}$ for the unpolarized cross section and below $10^{-6}$ for the angular distribution, rendering the asymptotic expressions sufficiently accurate for phenomenological applications.   

In phenomenology, the $\mathcal{O}(\alpha_s)$ correction (with $\mu_R=\sqrt{s}/2$) amounts to approximately $50\%$ of the LO cross section, while the $\mathcal{O}(v^2)$ and 
$\mathcal{O}(\alpha_s v^2)$ corrections are accidentally small, at $-1.9\%$ and $0.8\%$, respectively. The smallness of the relativistic corrections is attributed to the small SDCs associated with the $\mathcal{O}(v^2)$ matrix element. After including the feeddown contributions from $\psi(2S)$, the theoretical prediction for the cross section yields $0.523_{-0.197}^{+0.285}$ pb, which is consistent with the {\tt Belle} measurement within uncertainties. 

In contrast, the theoretical prediction for the $J/\psi$ angular distribution differs significantly from the experimental data. The predicted angular distribution parameter is $0.120_{-0.036}^{+0.041}$, which deviates from {\tt Belle} measurement
$5.71\pm 2.51$ by more than $2\sigma$, indicating the need for further experimental and theoretical investigations.

With the high luminosity anticipated at {\tt Belle} II, more precise measurements of $e^+e^-\to J/\psi+X$ are expected, which will help test the predictions of NRQCD factorization and deepen our understanding of the $J/\psi$ nature.

\acknowledgments

The work of C. L., H.-Y. L. and W.-L. S. is supported by the National Natural Science Foundation of China under Grants No. 12375079 and Fundamental Research Funds for the Central Universities under Grants No. SWU-KF26003. The work of X.-D. H. is supported by the National Natural Science Foundation of China under Grants No. 12505097, and the Chongqing Natural Science Foundation under Grant No. CSTB2025NSCQ-GPX1018.

\bibliography{refs.bib}

\end{document}